\documentclass{acm_proc_article-sp}
\usepackage{graphicx}
\usepackage{amssymb}
\usepackage{times}
\usepackage{multirow}
\usepackage{color}
\usepackage{verbatim}
\usepackage{amsmath}
\usepackage[vlined,english,ruled]{algorithm2e}
\usepackage{import}

\newcommand{\m}[1]{\mathcal{#1}}

\graphicspath{{figures/}}

\renewcommand{\S}[0]{{\mathcal S}}

\begin{document}
\conferenceinfo{GECCO'12,} {July 7-11, 2012, Philadelphia, Pennsylvania, USA.}
    \CopyrightYear{2012}
    \crdata{978-1-4503-1177-9/12/07}
    \clubpenalty=10000
    \widowpenalty = 10000
    
\title{On Neighborhood Tree Search}
\numberofauthors{2} 
\author{
\alignauthor
Houda Derbel\\
       \affaddr{FSEGS}\\
       \affaddr{Route de l'a\'eroport km 4, Sfax}\\
       \affaddr{Tunisia}\\
       \email{derbelhouda@yahoo.fr}
\alignauthor
Bilel Derbel\\
       \affaddr{Universit\'e Lille 1}\\
       \affaddr{LIFL CNRS -- INRIA Lille}\\
       \affaddr{France}\\
       \email{bilel.derbel@lifl.fr}
}

\date{}

\maketitle
\begin{abstract}
We consider the neighborhood tree induced by alternating the use of different neighborhood structures within a local search descent. We investigate the issue of designing a search strategy operating at the neighborhood tree level by exploring different paths of the tree in a heuristic way. We show that allowing the search to 'backtrack' to a previously visited solution and resuming the iterative variable neighborhood descent by 'pruning' the already explored neighborhood branches leads to the design of effective and efficient search heuristics. We describe this idea by discussing its basic design components within a generic algorithmic scheme and we propose some simple and intuitive strategies to guide the search when traversing the neighborhood tree. We conduct a thorough experimental analysis of this approach by considering two different problem domains, namely, the Total Weighted Tardiness Problem (SMTWTP), and the more sophisticated Location Routing Problem (LRP). We show that  independently of the considered domain, the approach is highly competitive. In particular, we show that using different branching and backtracking strategies when exploring the neighborhood tree allows us to achieve different trade-offs in terms of solution quality and computing cost.
\end{abstract}




\section{Introduction}
\label{sec:intro}

\textbf{Context and Motivation:} 
Metaheuristics are now considered as a well established algorithmic framework providing flexible and powerful tools to solve many hard optimization problems. 
Many efforts are being made by the research community in order to develop new search methods to help the design of both effective and efficient algorithms. 
In this paper, we build on previous techniques by developing an intuitive idea based on exploiting different neighborhoods in a forward-backward manner to explore what we term \emph{the neighborhood tree}. Generally speaking, we consider the possibility of making backward moves to a solution previously explored by some neighborhoods, and continue the search from there using other different neighborhoods searching for a better neighborhood combination. In the following, we first review some previous related works, then after, we give our contribution and describe our findings in more details.

\textbf{Background and related works:}
Among other search techniques, variable neighborhood search (VNS) and its several variants~\cite{VNS} are based on the systemic change of neighborhood within the search. For instance, Variable Neighborhood Descent (VND) exploits the idea of alternating between several neighborhoods within an iterative local improvement descent to escape local optima. More precisely, starting with a first neighborhood structure, VND performs local search until no further improvements are possible. From this local optimum, the local search is continued with the next neighborhood. If an improving solution is found, then the local search continues with the first neighborhood, otherwise the next available neighborhood is explored, and so on until no further improvements can be obtained. It is well known that the performance of VND can highly depend on the order the neighborhoods are alternated. In standard variants of VND, it is often admitted that ordering neighborhoods in an increasing cost/size is a reasonable strategy. However, this standard strategy is not always applicable, for instance, when the best ordering for a given problem can vary from one instance to another one. Actually, the issue of how to combine/exploit/search different neighborhoods is not new and one can find many different studies on the subject. For instance, in~\cite{PR}, a fast relaxation of neighborhoods is evaluated in order to select the most accurate ones. In~\cite{HR}, a self-adaptive strategy is used to rank neighborhoods and to dynamically choose the best suited ordering. A number of specific multi-neighborhood combination functions can also be found. For instance, many studies consider to take the union of some basic neighborhoods. The so-called neighborhood composition and the token-ring search are also other well known neighborhood combination functions, see e.g.,~\cite{LKG,GS,GMK,LGH}. More generally, hyperheuristics~\cite{Hyp} can be considered as a high level approach operating in the neighborhood space and aiming at producing effective hyper-search strategies. For instance, in~\cite{Cow,Ozcan}, simple hyperheuristic selection strategies are considered where low level heuristics (neighborhoods) are chosen either randomly, or greedily, or based on a score function. Several other sophisticated hyper-strategies, mainly inspired by the way metaheuristics operate, can be found in the literature, see e.g.,~\cite{Hyp2}.

\textbf{Technique overview and results:}
The study conducted in this paper is based on the simple observation that defining how neighborhoods are alternated within a local search descent is nothing other than defining a specific strategy to traverse a neighborhood tree, where the root of the tree represents the initial candidate solution and intermediate nodes represent solutions obtained by applying one of the possible neighborhoods. In other words, we view the trajectory of a variable neighborhood search as a high level neighborhood path, where path nodes are solutions and every path hop represents the exploration of one solution using one neighborhood among those available. Following this observation, we term \emph{a neighborhood tree search (NTS)} a strategy which is able to traverse the neighborhood tree efficiently searching for promising paths. It should be clear that a systematic traversal (exploration of all neighborhood branches) could not be efficient especially when the number of neighborhoods is high.

In this paper, we focus on the possibility of backtracking to previously visited solutions while branching and pruning tree nodes all along a search path. We show that this idea with basic iterative improvement descents leads to efficient search strategies both in terms of solution quality and computing cost. More specifically, we consider a simple randomized neighborhood selection strategy, where the choice of which neighborhood to select at runtime is made uniformly at random among those not yet explored. When effectively branching a neighborhood, we consider both deterministic and randomized adaptive strategies, basically relying on the neighborhood path traversed by the search in previous rounds. As for backtracking, we investigate intuitive strategies based on random and tournament selection techniques. We would like to emphasize that the proposed approach and its design components are \emph{generic} and \emph{not} specific to a fixed problem \emph{nor} to any particular neighborhood class.

We study the properties of the proposed approach by considering several instances coming from two different and well-studied problem domains: the Single Machine Total Weighted Tardiness Problem (SMTWTP) in the family of scheduling problems, and Location Routing Problem (LRP). Both problems are NP-Hard. Many previous studies have been successfully applied to solve the SMTWTP using hybrid variable neighborhood like searches. LRP is a more sophisticated problem which involves two simultaneous decisions: which depots to open and what routes to plan. Common to these two problems, many natural neighborhood structures can be considered making them two excellent case studies to analyze how our neighborhood tree based approach performs under different scenarios. Through extensive experiments, we show that our approach leads to substantial improvements in the solving of the two considered problems. More specifically, for SMTWTP we consider three neighborhood structures and we show that NTS performs better than standard VND executed with any neighborhood ordering, i.e., NTS is able to dynamically find its way along the neighborhood tree without any specific tuning. For LRP, we  consider eleven neighborhoods and a finely tuned VNS algorithm. Ultimately, we show that NTS is able to beat VNS without requiring any specific perturbation/shaking phase, but the backtracking it-slef. More importantly, VNS is used as a base-line algorithm allowing us to show how NTS performs when instantiating its components following different strategies. This allows us to give insights into the behavior of NTS and to better understand its critical design issues. In particular, we show that NTS can lead to different (and incomparable) trade-offs in terms of solution quality and running time. In a general point of view, our study reveals that NTS is a promising approach offering many interesting search abilities.\\
\textbf{Outline:} In Section~\ref{sec:nts}, we give an algorithmic scheme for NTS and describe intuitive strategies to be analyzed later. In Section~\ref{sec:prob}, we describe the considered problems and neighborhoods. In Section~\ref{sec:smtwtp} (resp. ~\ref{sec:lrp}), we analyze NTS and give our experimental results.

\section{Neighborhood tree search (NTS)}\label{sec:nts}

\subsection{Preliminaries}
Let us assume that we are given an optimization problem and a set of corresponding neighborhoods. We aim at designing an algorithm that exploits those neighborhoods as efficiently as possible. For simplicity, assume in addition that we have a \emph{step function} that given a candidate solution $s$ returns a solution $s'$ computed w.r.t. one neighborhood. Having an initial solution $s_0$, we then term the neighborhood tree $\m{T}$ the (possibly infinite) tree structure obtained by the following process. The root node of $\m{T}$ is the initial solution $s_0$. The first level of $\m{T}$, are the candidate solutions obtained from $s_0$ by applying the previously defined \emph{step function} w.r.t. every available neighborhood. The $j^{\textrm{th}}$ level is then constructed recursively from the internal nodes in the $(j-1)^{\textrm{th}}$ level and so on. Notice that this is a rather informal definition which is only given for the sake of illustration and to clarify our preliminary remarks.

It is clear that designing a local search algorithm exploiting the available neighborhoods can be viewed as designing a specific strategy to explore the considered neighborhood tree. This is what we term a neighborhood tree search (NTS) algorithm, i.e., a traversal strategy of the neighborhood tree searching for paths leading to promising regions. Designing such a traversal search algorithm can be difficult for many reasons. Firstly, for many optimization problems there may exist a relatively high number of natural neighborhood structures, say a constant $k>2$. Hence, a trivial exhaustive traversal of all tree nodes at height $g(n)$, a function of the problem size $n$, would require at least order of $k^{g(n)}$ steps, which can be intractable. Secondly, the goal is not to systematically traverse as many as possible tree nodes, but to find a path leading to high quality solutions while paying the minimum computing cost. Fortunately, we know that there exist heuristic traversal techniques leading to relatively efficient and effective search algorithms. For instance, this is the case for VND like search algorithms and many others that could be viewed as specific traversal strategies. In this paper, we focus on designing dynamic traversal heuristics where at each step, one have to decide what neighborhood to consider when going deeper in the neighborhood tree while backtracking to a previously visited solution whenever the search stacks into non promising tree regions.

\subsection{A basic randomized NTS}
Algorithm~\ref{algo:RNTS} gives a relatively detailed description of our first NTS example. As input, we assume that we are given a set of $k$ neighborhood structures $\{\m{N}_1, \cdots, \m{N}_k\}$ relative to a given problem and a fitness/evaluation function $f$ to be minimized. Algorithm~\ref{algo:RNTS} maintains a trajectory path (variable $Path$) containing an ordered sequence of visited solutions with their neighborhood usage. This is encoded by variable $h_s=(h_s^1, \cdots, h_s^k)$ where $h_s^i$ is $1$ whenever neighborhood $\m{N}_i$ has been used to explore solution $s$ and $0$ otherwise. The algorithm then proceeds iteratively by considering the last solution $s$ (function $\textsc{Head}$) appearing in the trajectory path. For that solution, a neighborhood $\m{N}_i$ is chosen uniformly at random among those that have not been used to explore $s$. A $\textsc{Step}$ function is then applied to compute a new solution $s'$ and neighborhood usage variable $h_s^i$ is updated. The $\textsc{Step}$ function could be for instance any local search heuristic based on neighborhood structure $\m{N}_i$, e.g., a hill-climbing. Having explored a new neighborhood, we shall decide whether to continue the search with solution $s'$ or to backtrack. In Algorithm~\ref{algo:RNTS}, a simple acceptance criterion is used. More precisely, if the new explored solution $s'$ is found to improve $s$ then, we make a move forward by pushing $s'$ at the end the trajectory path. Otherwise, we check whether there exists some neighborhoods which have \emph{not} been used to explore $s$. If such a situation exists, we simply continue the inner-loop, that is we try to find an improving solution by selecting uniformly at random a non explored neighborhood w.r.t. the current solution $s$. Otherwise, a backtrack move is activated ('else' condition). Notice that in this case, current solution $s$ is not necessarily a local optimum w.r.t. all neighborhoods. In fact, this depends on the $\textsc{Step}$ function and the depth of $s$ in the search trajectory. Backtracking in Algorithm~\ref{algo:RNTS} is done using a simple uniform randomized selection process. More precisely, among path solutions which are not yet explored by all neighborhoods, one is chosen uniformly at random, say solution $s_j$ at position $j$. The trajectory path is then updated by deleting those solutions laying between $s_j$ and $s$. The search then continues from $s_j$ in the same way until the trajectory path becomes empty.

\begin{algorithm}
\KwIn{A set of $k$ neighborhood structures $\{\m{N}_1, \cdots, \m{N}_k\}$}
$s \leftarrow $ initial solution \;
$(h_s^1, \cdots, h_s^k) \leftarrow (0, \cdots, 0)$ ;
 $Path \leftarrow ~\big\{(s, (h_s^1, \cdots, h_s^k))\big\}$ \;
 
\Repeat{$Path = \emptyset$}{
	\CommentSty{/$^{**}$ Current trajectory solution $^{**}$/}\\
	$(s,h_s) \leftarrow \textsc{Head}(Path)$ \;
	\CommentSty{/$^{**}$ Neighborhood Selection $^{**}$/}\\
	$I_s \leftarrow \{ \ell ~~|~~ h_s^\ell = 0\}$ \;
	$i \leftarrow \textsc{Random}(I_s)$ \;
	
	\CommentSty{/$^{**}$ Neighborhood Exploration $^{**}$/}\\
	$s' \leftarrow \textsc{Step}(s,\m{N}_i)$\;
	$h_s^i \leftarrow 1$ \; 

	\CommentSty{/$^{**}$ Move or Backtrack $^{**}$/}\\
	\uIf{$f(s) < f(s')$}{
		$h_{s'} \leftarrow (0, \cdots, 0)$ \;
		$Path \leftarrow ~\textsc{Push}\big((s', h_{s'}), Path\big)$ \;
	}\ElseIf{$|I_s| + 1  = k$}{
		$Path \leftarrow \textsc{Random\_Backtrack}(Path)$\;
	}
} 
\caption{A simple randomized variant of NTS}
\label{algo:RNTS}
\end{algorithm}

\subsection{NTS generic scheme and variants}
Algorithm~\ref{algo:RNTS} given in the previous section is clearly a specific implementation of the more generic scheme given in Algorithm~\ref{algo:NTS}. Generally speaking, Algorithm~\ref{algo:NTS} is in fact an attempt to give a high level procedural description of a NTS like algorithm. In particular, we identify the \emph{history} variable which is typically used to record information about search trajectory and neighborhood performance. We also have the neighborhood selection stage which role is to help the search going towards promising neighborhood branches. The $\textsc{Step}$ and $\textsc{Accept}$ function, play the role of branching/pruning. These two functions should be thought in the same way classical local search algorithms operate, but keeping in mind that possibly several neighborhoods can be used. The third main stage of Algorithm~\ref{algo:NTS} is backtracking. Within NTS, backtracking serves mainly to adapt the traversal when it is stack into paths that do not lead to improvements. Backtracking should be thought with respect to search history. In this paper, we study the properties of NTS by considering the following particular variants.

\begin{algorithm}
\KwIn{A set of $k$ neighborhood structures $\{\m{N}_1, \cdots, \m{N}_k\}$.}
$s \leftarrow $ initial solution; $history \leftarrow \varnothing $\;

 
\Repeat{\textsc{stopping\_condition}}{
	\CommentSty{/$^{**}$ Neighborhood Selection Strategy $^{**}$/}\\
	$i \leftarrow \textsc{Select}(s,history)$ \;
	\CommentSty{/$^{**}$ Branching/Pruning Strategy $^{**}$/}\\
	$s' \leftarrow \textsc{Step}(s,\m{N}_i,history)$\;
	\eIf{$\textsc{Accept}(s,s',history)$}{
		$s \leftarrow s'$ \; 
	}{
		\CommentSty{/$^{**}$ Backtracking Strategy $^{**}$/}\\
		$s \leftarrow \textsc{Backtrack}(history)$ \;
	}	
} 
\caption{general purpose design scheme for NTS}
\label{algo:NTS}
\end{algorithm}

\textbf{Trajectory history}: In all our variants, we record the path traversed by the search and containing the branching solutions and their relative neighborhood usage (exactly in the same way than in Algorithm~\ref{algo:RNTS}). We additionally record the (local) best fitness $f_s^{best}$ relative to each solution $s$ and observed by the search when the local step function $\textsc{Step}(s,\m{N}_i)$ is applied at position $s$.

\textbf{Neighborhood Selection Strategy}: We simply consider the randomized process depicted in Algorithm~\ref{algo:RNTS}, i.e., one neighborhood among those not previously used at current path position is selected uniformly at random.

\textbf{Step function}  $\textsc{Step}(s,\m{N}_i)$ : we study four classical alternatives denoted BI (best improvement), FI (first impr.), BD (best descent) and FD (first descent). BI corresponds to the case where solution $s'$ is the neighbor of $s$ (w.r.t $\m{N}_i$) with the best fitness. For strategy FI, $s'$ is the first neighbor of $s$ which is found to have a better fitness than $s$, when processing $s$ neighbors in a random order. BD (resp. FD) denotes a local search descent where BI (resp. FI) strategy is applied until no improving neighbors can be found.

\textbf{Acceptance/Branching criterion}: We study three strategies denoted AA, AI, and AT. AA is exactly the same than in Algorithm~\ref{algo:RNTS}, i.e., any solution $s'$ that improves the fitness of current solution $s$ is accepted : $f(s') < f(s)$. AI denotes the strategy where solution $s'$ is accepted if its fitness $s$ is better than $f_s^{\textrm{best}}$, i.e., $f(s') < f_s^{\textrm{best}}$. AT is a combination of AA and AI. More specifically, a solution $s'$ is always accepted if $f(s') < f_s^{\textrm{best}}$. Otherwise, if $f(s') > f_s^{\textrm{best}}$, but $f(s') < f(s)$ then $s'$ is accepted with a probability parameter $p_a$. Otherwise $s'$ is not accepted. In our study, we adopt an adaptive strategy where $p_a=1/d(s)$ with $d(s)$ the position of solution $s$ in the trajectory path. In other words, a neighborhood, leading to a branch improving  $s$, but not improving the previous best local fitness obtained using a different neighborhood, is accepted with a probability which is proportional to the branch height in the neighborhood tree, i.e., the more we are deep in the neighborhood tree, the more it is unlikely to explore non improving branches.

\textbf{Backtracking strategy}: We consider three backtracking strategies denoted BR, BH, BU. BR is the strategy depicted in Algorithm~\ref{algo:RNTS}, i.e., among path positions which are not yet explored by all neighborhoods, one is chosen uniformly at random. BH and BR are more sophisticated tournament-based selection strategies. More precisely, for both BH and BR, we select two distinct path positions $s_j$ and $s_{j'}$ uniformly at random among those not yet explored by all neighborhoods. With BH, we backtrack to the solution which is less deep in the trajectory path, i.e., if $s_{j'}$ appears after $s_{j}$ in the search path, then we backtrack to $s_j$. With BU, we backtrack to the solution which was explored less often by available neighborhoods.

\textbf{Stopping Condition}: We consider two different stopping conditions: we end the search when (i) the path trajectory is empty, or (ii) a maximum number of fitness evaluations is reached. 

\textbf{Terminology and notations:} For clarity, we shall use the following notation \textbf{NTS-$\mathbf{(X,Y,Z)}$} where $\mathbf{X\in\{FI,BI,FD,BD\}}$ is the step function, $\mathbf{Y\in\{AA,AI,AT\}}$ the branching/acceptance strategy, and $\mathbf{Z\in\{BR,BH,BU\}}$ is the backtracking strategy.

\section{Problem domains}\label{sec:prob}
\subsection{Single machine scheduling (SMTWTP)}
\textbf{Problem definition and motivation}: In the Single Machine Total Weighted Tardiness Problem, we are given n jobs. Each job has to be processed without any interruption on a single machine that can only process one job at a time. Each job has a processing time $p_j$ , a due date $d_j$ and an associated weight $w_j$ (reflecting the importance of the job). The tardiness of a job $j$ is defined as $T_j = \max \{0, C_j - d_j \}$, where $C_j$ is the completion time of job $j$ in the current sequence of jobs. The goal is then to find a job sequence minimizing the sum of the so-called weighted tardiness: $\sum_{i=1}^{n} w_i \cdot T_i$. SMTWTP is NP-hard. Several different metaheuristics have been proved to efficiently solve SMTWTP benchmark instances, e.g.,~\cite{vnd,ILS,dyna} to cite a few. SMTWTP is in fact a well understood problem which is often used to study the properties of search heuristic methods. This paper is not an exception. Although we are able to show that very simple NTS techniques outperform previous more sophisticated and finely tuned heuristics for SMTWTP, we shall rather focus on studying and understanding the behavior of our NTS heuristics.\\
\textbf{Neighborhoods}: Permutations are the standard representation used for SMTWTP. In this paper, we consider three standard neighborhoods. \textit{Exchange (E)}: all permutations that can be obtained by swapping adjacent jobs in the permutation. \textit{Swap (S)}:  all permutations that can be obtained by swapping adjacent jobs at the $i^{\textrm{th}}$ and $j^{\textrm{th}}$ position. \textit{Insert (I)}: all permutations that can be obtained by removing a job at position $i$ and inserting it at position $j$.\\
\textbf{Instances}: We consider the well known $100$ Job instance set, and at a less extent, the $50$ and $40$ Job instances from the OR-Library~\cite{or}. Each instance set contains $125$ instances.

\subsection{Location Routing problem (LRP)}
\textbf{Problem definition and motivation}: LRP~\cite{lrp,Min} deals with two NP-hard problems, namely, facility location problem (FLP) and vehicle routing problem (VRP). Roughly speaking, in LRP one has to simultaneously decide which depots to open and what routes to establish to satisfy client demands. 
Besides being a challenging problem for local search heuristics, LRP is of special interest since many neighborhoods can be naturally considered for both the location and the routing level (which are known to be inter-dependent). Many specific search strategies have been studied for LRP, e.g.,~\cite{galrp,lrpvns,duhamel} to cite a few. A common aspect in these studies is to find a good balance for simultaneously searching the routing level and the location level. In particular, there exist a rich literature on several different neighborhoods dealing with the two LRP levels. This makes the choice and the combination of neighborhoods critical and thus LRP is an excellent candidate problem to study our approach. In this paper, we consider the uncapacitated vehicles and capacitated LRP~\cite{ulrp}. More specifically, we consider a set of $n$ customers and a set of $m$ potential depots. Each depot has a limited capacity and a fixed opening cost. Each depot is associated with a single uncapacitated vehicle. Each customer has a non-negative demand which is known in advance and should be satisfied. For any pair of clients (or client-depot), there is an associated traveling cost. LRP then consists in minimizing the total cumulative cost of both depot opening (location) and client delivery (Routing).\\
\textbf{Neighborhoods}: We consider a natural representation of a candidate solution (not necessarily feasible) for LRP, namely, a list of opened and non opened depots. For each depot, a permutation represents the assigned clients and their order in the route. We consider eleven neighborhoods sketched in the following. $\m{N}_1$ (resp. $\m{N}_2$): All solutions obtained by performing a client \emph{insertion} move in the permutation(s) encoding one single depot route (resp. two \emph{different} depot routes). $\m{N}_3$ (resp. $\m{N}_4$): All solutions obtained by performing a swap move on the permutation(s) encoding one depot route (resp. two depot routes). $\m{N}_5$ (resp. $\m{N}_6$): All solutions obtained by performing a classical $2$-opt move on the permutation(s) encoding one depot route (resp. two depot routes). $\m{N}_7$ (resp. $\m{N}_8$): All solutions obtained by performing an extended insertion move on the permutation(s) encoding every route (resp. two \emph{different} routes), that is an insertion of any sub-route of any possible length, i.e., route bone insertion. $\m{N}_{9}$ (resp. $\m{N}_{10}$): All solutions obtained by performing a route bone insertion move as for neighborhoods $\m{N}_7$ (resp. $\m{N}_8$) but while inserting clients sub-route in the reverse order. $\m{N}_{11}$: All candidate solutions obtained by closing a depot and affecting its whole route to a closed depot (close one depot and open a new one).\\
Since we are dealing with neighborhoods producing possibly unfeasible solutions, we use an evaluation function $f$ defined as following: $f(s)=c(s)+p(s)$ where $c(s)$ is the cost of $s$ as stated by the objective function of LRP and $p(s)$ is a penality on the violation of depot capacity constraints. It is calculated by the equation: $p(x) = \sum_{j}  \alpha \cdot \max\{0, Q_j (s) - b_j \}$ where $\alpha$ is a weight factor parameter, $Q_j (s)$ is the total demand of customers serviced by depot $j$ and $b_j$ is the capacity associated with depot $j$, i.e., the more depot constraints are violated, the more is the penality and the more the search is forced to move toward feasible regions.\\
\textbf{Instances}: We consider a set of $450$ instances taken from the literature~\cite{ulrp}. These instances can be grouped into finely defined classes according LRP specific parameters. In our experimental results, we simply group them into $5$ sets according to the number of clients ($n$) and the number of depots ($m$): $(n,m) \in \{(5,10),$ $(5,20), (5,30), (10,20), (10,30)\}$. Each set is containing equally the same number of instances, i.e.,~$90$.

\section{Experimental analysis: SMTWTP}\label{sec:smtwtp}

\begin{table*}[htb!]
\caption{Results for standard VNDs with a random initial solution. Results are for the $100$ job instances with $30$ trials per instance. First column gives neighborhood ordering, i.e., Exchange (E), Insert (I), Swap (S). FI, BI, FD and BD columns are for the local step functions given in Section~\ref{sec:nts}. $N_{\textrm{opt}}$ is the number of optimal solutions found by at least one trial. $\overline{\Delta}$ is the average percentage deviation from optimal and $\overline{\textrm{Eval}}$ is the average number of evaluations until search termination. Bold style is for best result (for each column).
\vspace{-3ex}
}
\label{tab:vnd}
$$
\begin{array}{|c|||c|c|c||c|c|c|||c|c|c||c|c|c|||}
\hline
\multirow{2}{*}{\textrm{Order}} & \multicolumn{3}{|c||}{\textrm{FI}} & \multicolumn{3}{|c|||}{\textrm{BI}}  & \multicolumn{3}{|c||}{\textrm{FD}} & \multicolumn{3}{|c|||}{\textrm{BD}} \\ \cline{2-13}
 \\[-9 pt] 
\cline{2-13}
& N_{\textrm{opt}} & \overline{\textrm{Eval}}  \rule[-5pt]{0pt}{15pt} & \overline{\Delta} & N_{\textrm{opt}} & \overline{\textrm{Eval}} &  \overline{\Delta} & N_{\textrm{opt}} & \overline{\textrm{Eval}}  \rule[-5pt]{0pt}{15pt} & \overline{\Delta} & N_{\textrm{opt}} & \overline{\textrm{Eval}} &  \overline{\Delta} \\ \hline

\textrm{EIS} &102 & 140872.4 & \mathbf{0.010} & \mathbf{103} & \mathbf{471156.4} & \mathbf{0.015} & \mathbf{113} & 121787.3 & 0.015 & 101 & \mathbf{667976.4} & 0.019  \\ \hline

\textrm{ESI} & \mathbf{106} & 121372.3 & 0.016 & 91 & 485211.7 & 0.018 & 102 & 106214.4 & 0.015 & 92 & 946366.1 & 0.019 \\ \hline

\textrm{IES} & 104 & \mathbf{120262.4} & 0.017 & 96 & 1024409.3 & 0.022 & 95 & 101053.6 & 0.017 & 93 & 1006583.9 & 0.021 \\ \hline

\textrm{ISE} & 101 & 135572.1 & 0.013 & 96 & 865525.5 & 0.022 & 112 & 115144.1 & \mathbf{0.014} & 103 & 871254.1 & \mathbf{0.018} \\ \hline

\textrm{SEI} & 98 & 121220.9 & 0.017 & 93 & 1026960.0 & 0.021 & 102 & \mathbf{100903.9} & 0.016 & 92 & 1007357.2 & 0.025\\ \hline

\textrm{SIE} & \mathbf{106} & 138007.4 & 0.015 & 101 & 864900.3 & 0.023  & 106 & 114412.0 & \mathbf{0.014} & \mathbf{105} & 871631.1 & 0.021 \\ \hline
\end{array}
$$
\end{table*}

\subsection{Results with standard VND}
For SMTWTP, we consider to study the behavior of our approach compared to standard VND techniques. In Table~\ref{tab:vnd}, we report a summary of results we have obtained for SMTWTP when running a standard VND using the $6$ possible ordering of neighborhoods and the $4$ possible strategies for making a local step move. In accordance with results reported in previous studies, e.g.,~\cite{vnd}, VND is well suited for solving SMTWTP. In fact, the average percentage deviation from optimal is relatively low and around $90\%$ of instances are solved to optimality by at least one trial over $30$ performed in our experiments. However, one can notice that some instances remain hard to solve by standard VND as reported in previous works. In addition, no fixed ordering nor local step strategy outperforms all the others for all three measures reported in Table~\ref{tab:vnd}.

\subsection{Results overview with NTS}
\begin{figure}[htb!]
\begin{center}
\begin{tabular}{cc}
\hspace{-9ex}\includegraphics[width=0.3\textwidth]{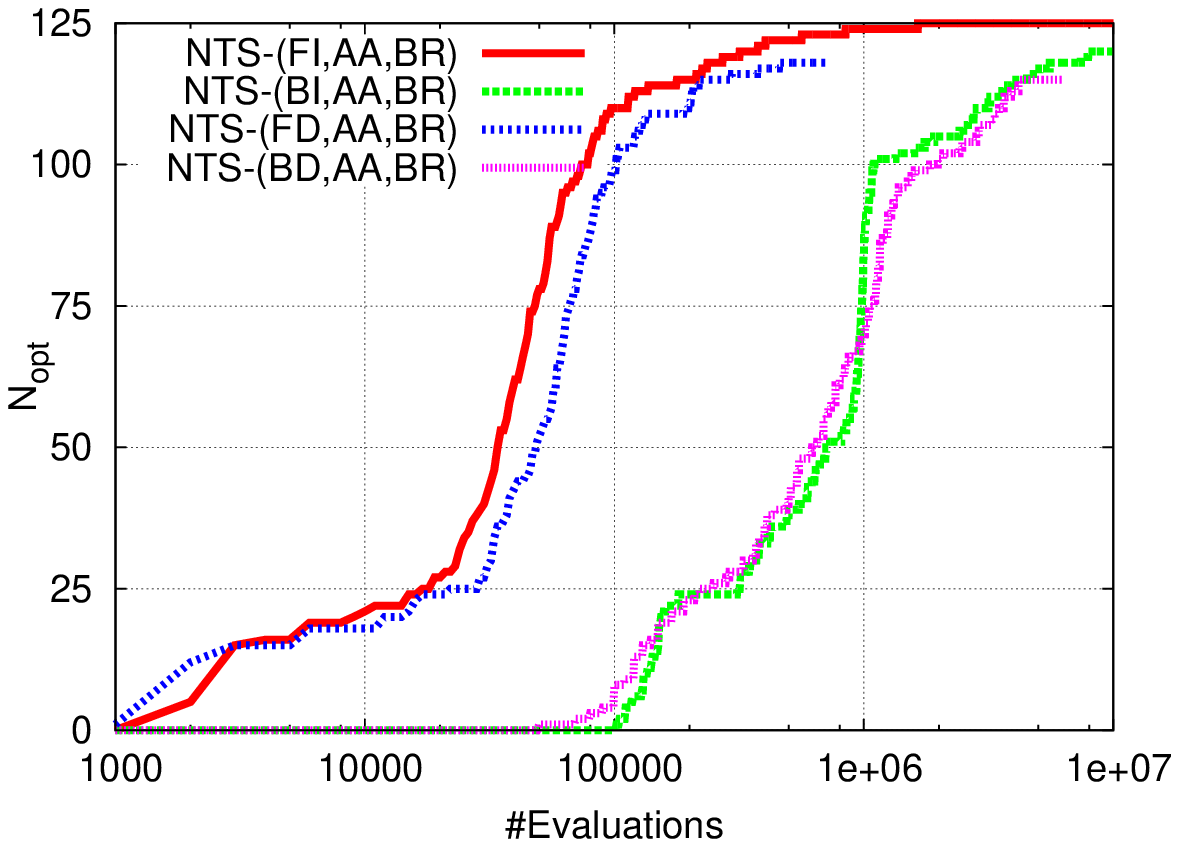} & \hspace{-5ex}\includegraphics[width=0.3\textwidth]{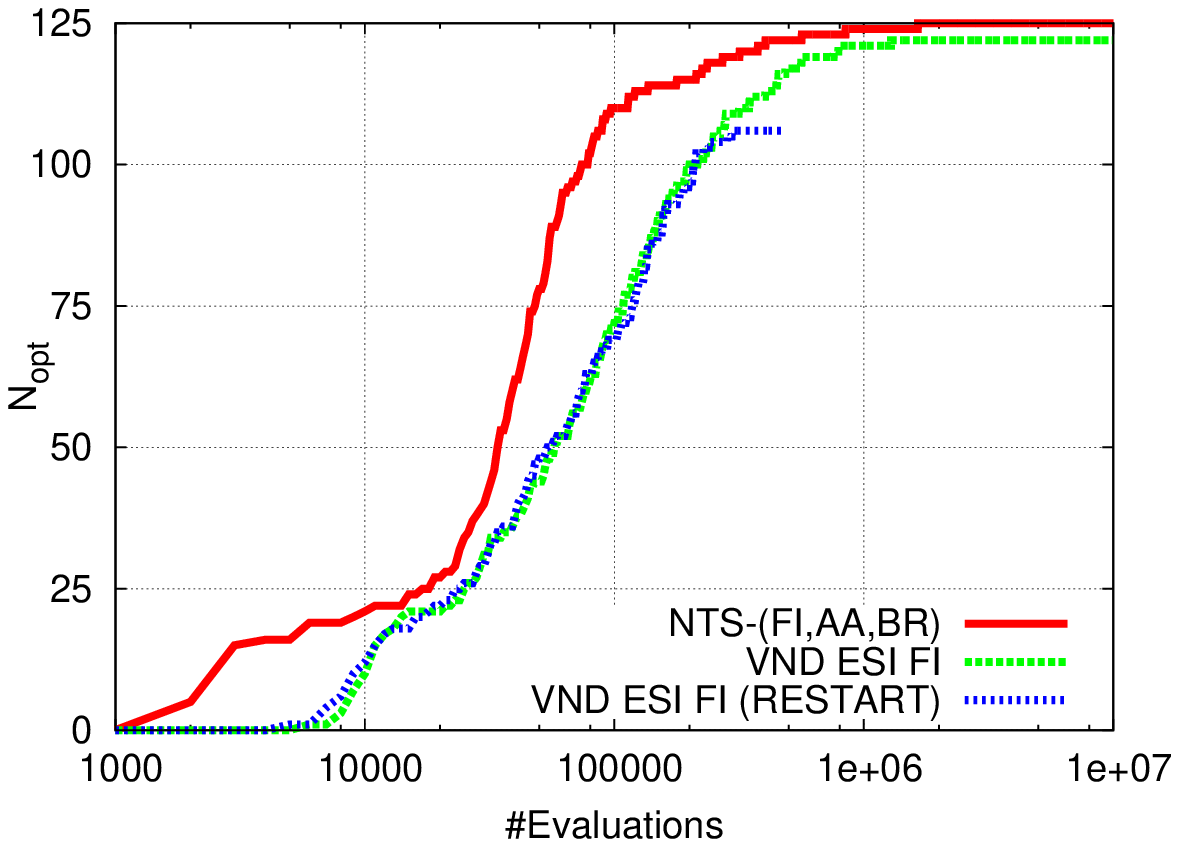}\\
\hspace{-9ex}\includegraphics[width=0.3\textwidth]{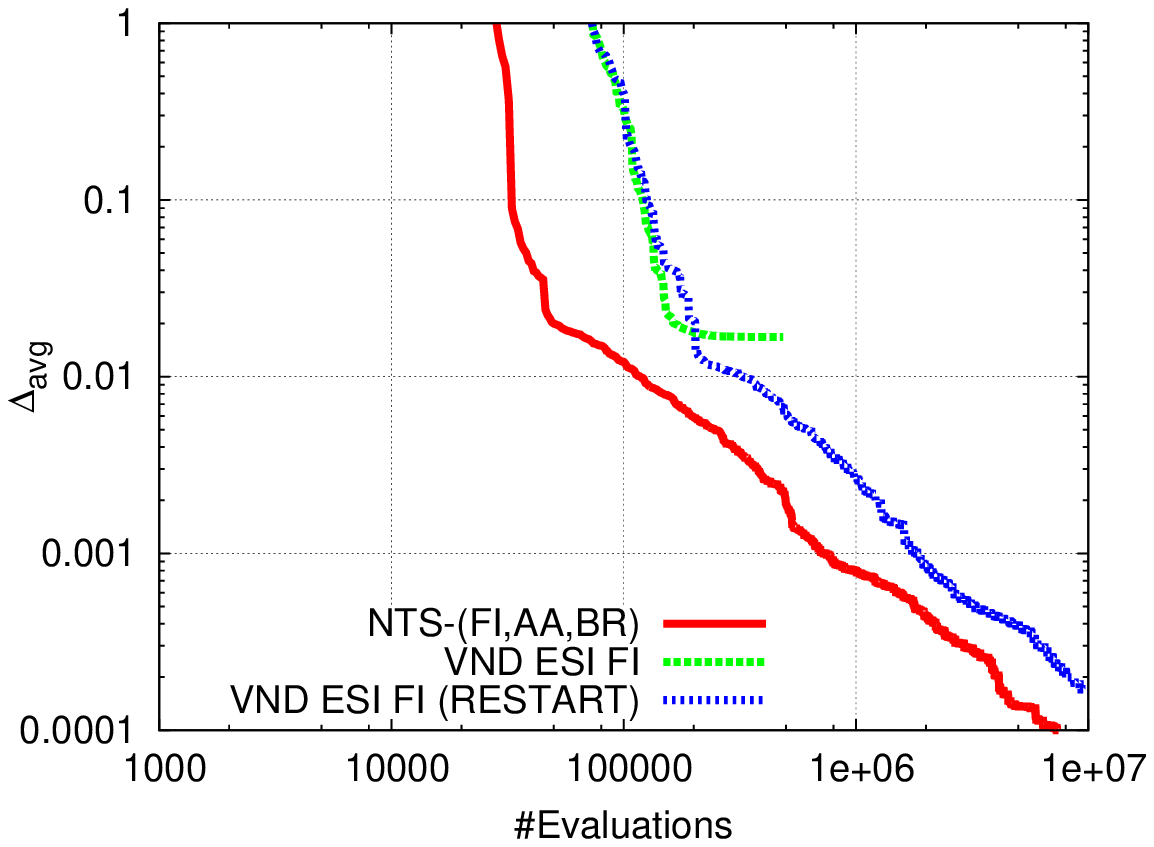} & \hspace{-5ex}\includegraphics[width=0.3\textwidth]{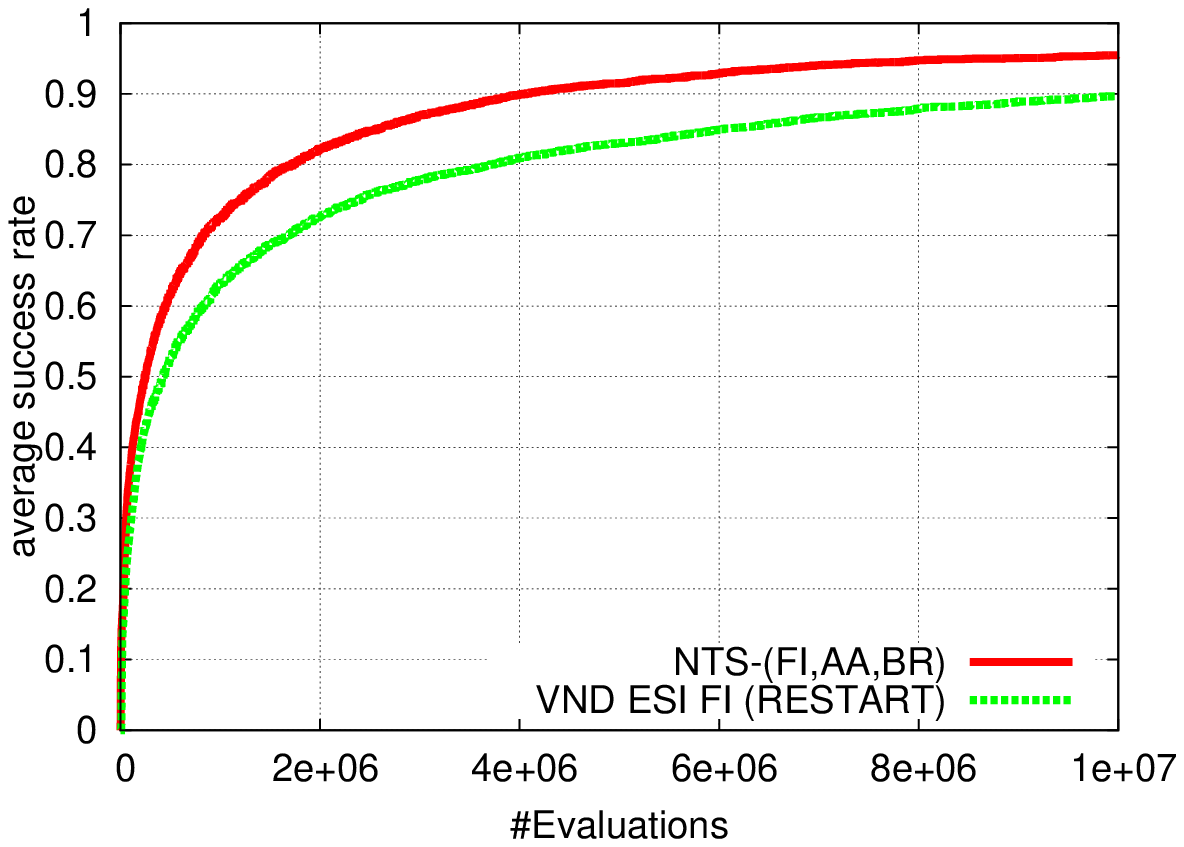}\\
\end{tabular}
\vspace{-4ex}
\caption{Results for NTS-(*,AA,BR) Vs standard VND for $100$ job instances (labels refer to search strategies). Top: cumulative number of instances solved to optimality ($N_{\textrm{opt}}$) by at least one trial over $30$  as a function of number of fitness evaluations. Bottom-Left: Evolution, with number of evaluations, of the average percentage deviation from optimal ($\overline{\Delta}$) averaged over $30$ trials. Bottom-Right: Evolution, with number of evaluations, of the success rate averaged over the $125$ instances.}
\label{fig:smtwtp_best}
\end{center}
\end{figure}

For SMTWTP, the reported results are obtained with the simple variant of NTS given in Algorithm~\ref{algo:RNTS}, with random initial solution, acceptance strategy AA, and backtracking strategy BR, i.e., NTS-$(*,AA,BR)$. As stopping condition, the search terminates when either a maximum number of evaluations, namely $10^7$, is reached, or the trajectory path becomes empty. Our results for NTS are summarized in Fig.~\ref{fig:smtwtp_best}. Firstly, we remark that for both FD and BD local step strategies, the search terminates before the maximum number of evaluations is reached (Fig.~\ref{fig:smtwtp_best} Top-left). At the opposite, for both FI and BI the search continues without the backtracking being able to force termination. This is mainly due to the relatively high trajectory path length as we will discuss later. Furthermore, as could be expected, first improvement strategies are less costly compared to best improvement strategies. We also found that FI outperforms all other step strategies both in computing cost and number of instances solved to optimality. In fact, it is the only strategy which is able to find an optimal solution (over the $30$ performed trials) for all the $125$ instances.\\
Moreover, NTS using step functions FI and FD can be proved to provide substantial improvements in all aspects over all the standard VND variants reported in Table~\ref{tab:vnd}. For the sake of clarity, we only report our finding using one VND ordering, namely ESI. Notice however that similar conclusions can be drawn for the other possible orderings. To be fair in our comparative study, we further consider restarting the VND algorithm from a randomly generated solution in the case VND terminates before the maximum number of evaluations is reached. As shown in Fig.\ref{fig:smtwtp_best} (Top-right and Bottom), NTS outperforms VND both in terms of: cumulative number of instances solved to optimality ($N_{\textrm{opt}}$), average percentage deviation from optimal ($\overline{\Delta}$), and average success rate that is the percentage of trials that do find the optimal solution, i.e., this can be interpreted as the probability distribution of finding the optimal solutions for all instances.

\subsection{Run Time Distribution Analysis}
\begin{figure}[htb!]
\begin{center}
\begin{tabular}{cc}
\hspace{-9ex}\includegraphics[width=0.3\textwidth]{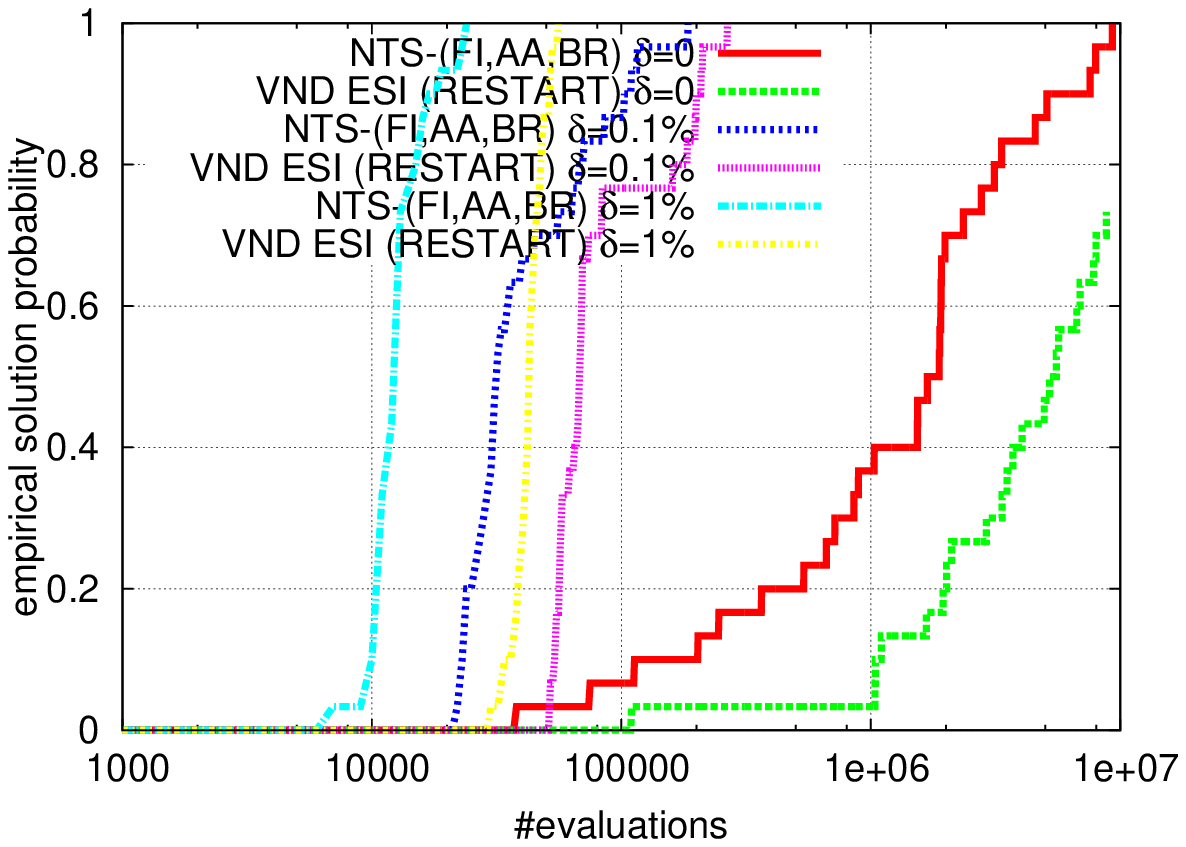} & \hspace{-5ex}\includegraphics[width=0.3\textwidth]{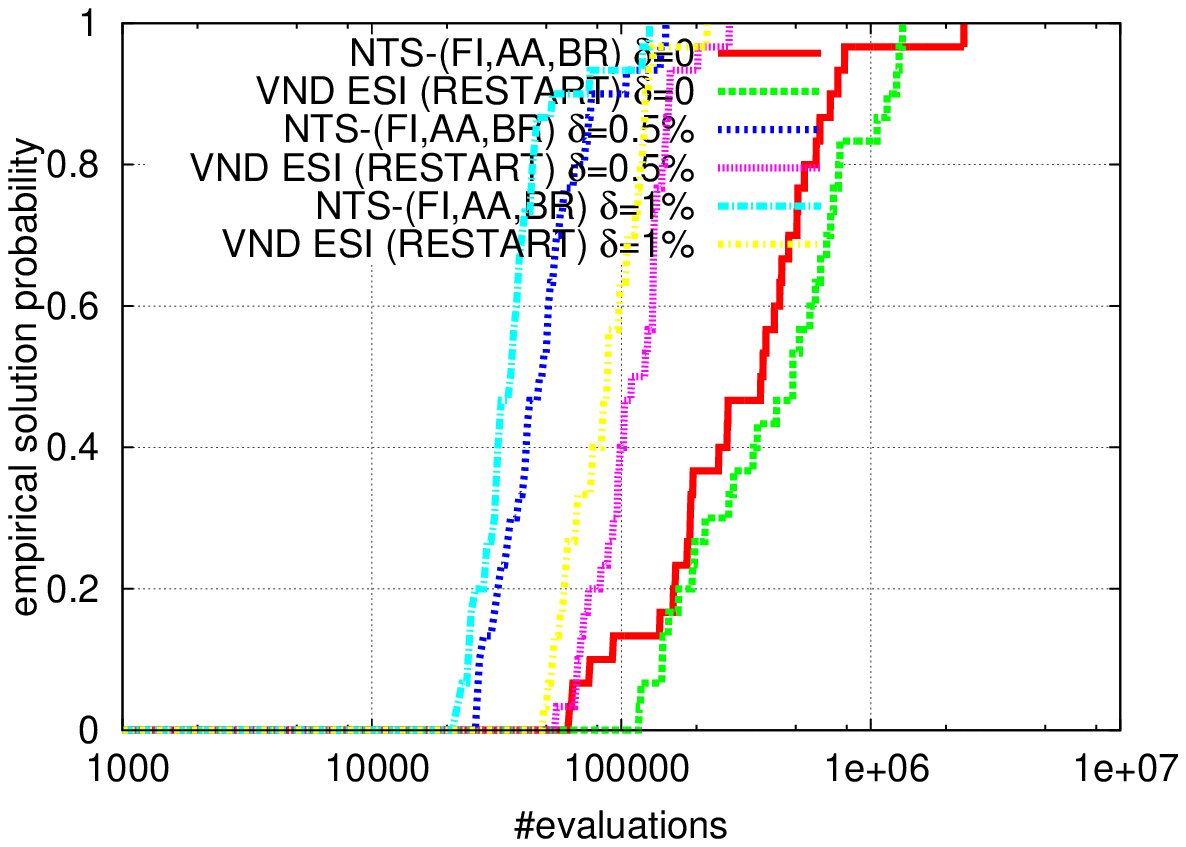}\\
\hspace{-9ex}\includegraphics[width=0.3\textwidth]{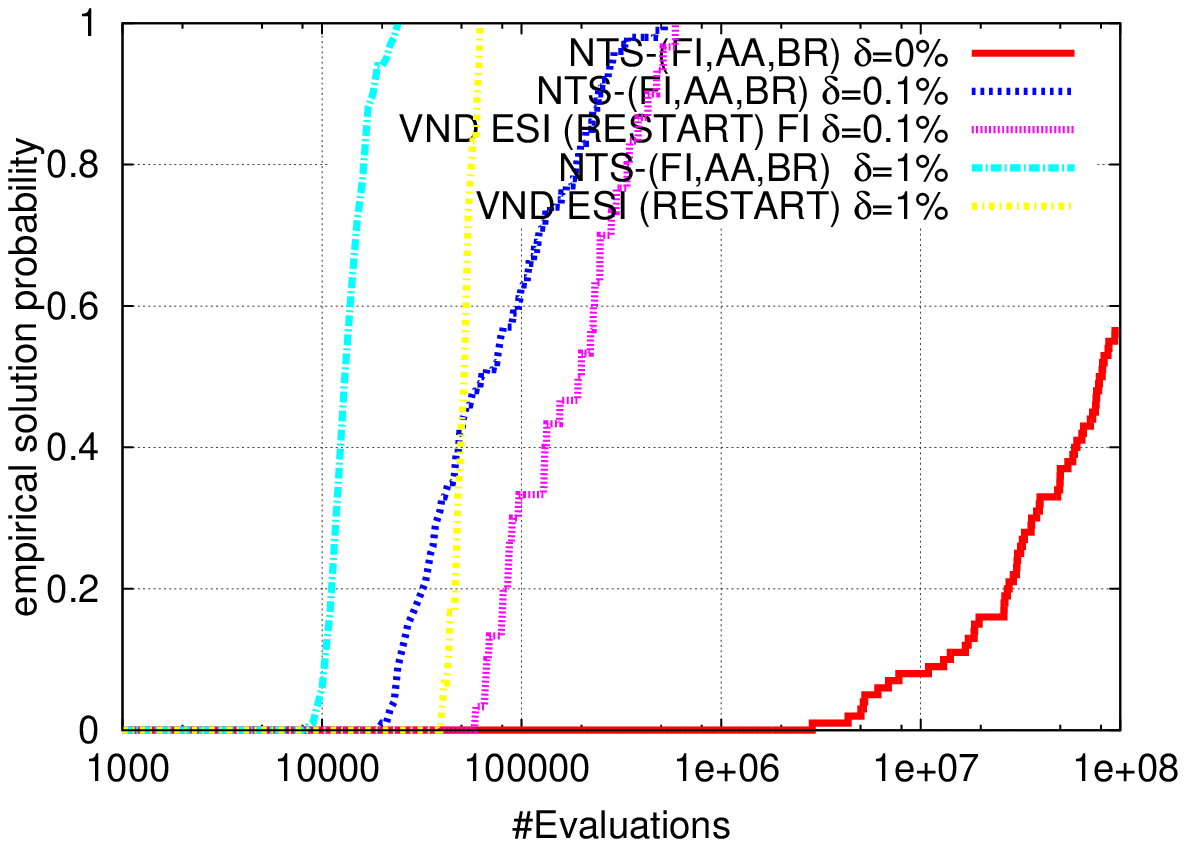} & \hspace{-5ex}\includegraphics[width=0.3\textwidth]{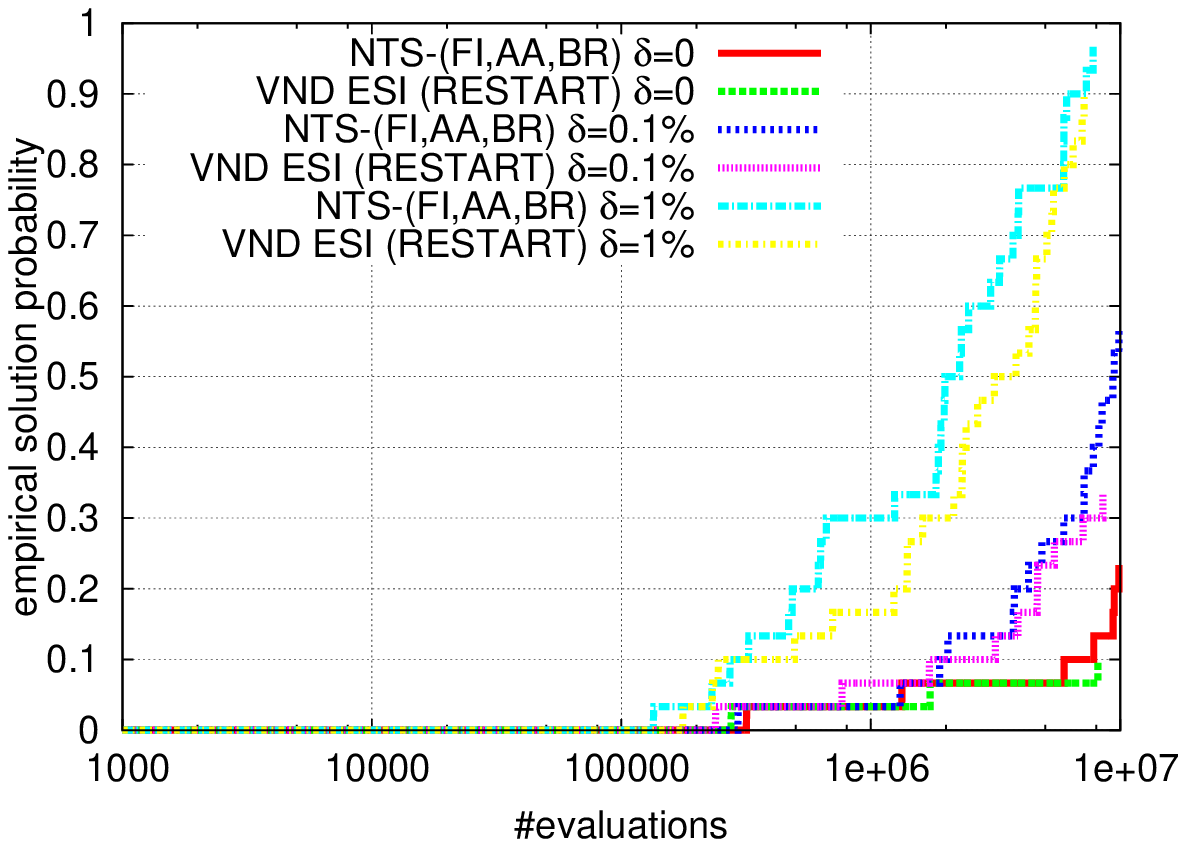}\\
\end{tabular}
\vspace{-4ex}
\caption{RTDs for NTS \textit{vs} VND with random restart. The x-axis gives
the logarithm of the number of fitness evaluations, the y-axis the cumulative empirical solution quality probability. $\delta$ denotes the gap (in percentage) between the required solution quality and the optimal solution. Top-Left figure (resp. Top-Right, Bottom-Left and Bottom-Right) shows the results for instance 19 (resp. 38, 42, and 86).}
\label{fig:rtd}
\end{center}
\end{figure}

In previous section, we showed that NTS performs better than VND in general, i.e., results are mainly averaged over instances. In this section, we go to a throughout comparative study. More specifically, we analyze the run-time behavior of NTS compared to standard VND by using run-time distributions (RTD)~\cite{HS}. RTDs give the cumulative empirically observed probability of finding an optimal solution (or a solution within a specific quality bound) for a given instance as a function of the CPU time. In our study, we use a slightly different definition, where probability is considered as a function of number of evaluations. This is mainly  to stay independent of any specific implementation or operating system issues (NTS running time issues are however studied in next section). We examined the behavior of NTS with different step functions and for several instances, mainly, those who are reputed to be relatively hard. In Fig.~\ref{fig:rtd} we present our results for only four instances, but similar conclusions can be made for the others. The RTDs clearly shows that NTS performs better than VND for three out of the four instances, namely, $19$, $38$, and $41$, even if optimality is not required. For instance $86$, the difference is less pronounced, with a small advantage in favor of NTS for fitness deviation of $1\%$ from optimal.

\subsection{NTS history analysis}
Here, we give some basic observations about trajectory path used by NTS. In Table~\ref{tab:height}, we report the maximum length $h_{max}$ of the trajectory path ever observed for any instance and any trial , and $\overline{h_{max}}$ the maximum length ($30$ trials per one instance) averaged over $125$ instances of each problem size set.

With the FD step function, the maximum length for the three instance sets is at most $8$ which is relatively very low. This is a crucial observation that can be exploited in different manners for backtracking. Let $h$ be the maximum length that can be observed which is also the maximum height for the neighborhood tree. Assuming that the exploration of each of the $k$ neighborhoods by the step function FD requires a polynomial time in the problem size, say $p(n)$, then an exhaustive neighborhood tree traversal would lead to an NTS algorithm with computing complexity of roughly $O(k^h \cdot p(n))$. If $h$ is proved to be a constant or even a very small function in $n$, then an exhaustive neighborhood tree traversal could lead to a relatively efficient NTS. In our experiments, $h$ seems to be very insensitive to $n$ which suggests that other backtracking strategies, e.g., using a fixed/adaptive number of backtrack steps, could improve the search.

For the FI step function, it is clear that (compared to FD) there is a significant increase in the trajectory path length. Therefore, the previous discussion does not hold anymore using FI. In fact, assuming that $h_{\max}=\Omega(n)$, which seems to be the case, an exhaustive search of the neighborhood tree is normally intractable. This is further confirmed by the fact that in all our experiments with the FI strategy, NTS always reaches the maximum number of iterations without emptying the whole search path. We however remark that using random backtracking (BR), FI produces better results than FD. We attribute this to the fact that FI implies a neighborhood tree of relatively high size but relatively diversified neighborhood paths.

To conclude this section, we would like to give some remarks on the impact of maintaing and accessing the history path on CPU running time. As discussed before, path trajectory is relatively low compared to problem size. Knowing that the size of many neighborhood structures is at least linear, and many often polynomial, in problem size, the cost of maintaining the path history stays relatively marginal. In our implementation using standard Java library, without any specific code optimization, the average CPU-time to perform one evaluation for problem size $100$ on a standard 2.0 Ghz Intel processor is $0.0005$ millisecond.
\begin{table}[htb!]
\caption{NTS maximum path length ($\textrm{h}_{\max}$ and $\overline{\textrm{h}_{\max}}$)}
\label{tab:height}
\vspace{-3ex}
$$
\begin{array}{|c||c|c||c|c||}
\hline
\multirow{3}{*}{\textrm{n}} & \multicolumn{2}{|c||}{\textrm{FI}} & \multicolumn{2}{|c||}{\textrm{FD}} \\ \cline{2-5}
 \\[-9 pt] 
\cline{2-5}
& \textrm{h}_{\max} & \overline{\textrm{h}_{\textrm{max}}} \rule[-5pt]{0pt}{15pt}  &  \textrm{h}_{\max} & \overline{\textrm{h}_{\max}}  \\ \hline

40 & 267 & 166.3 &  7 & 3.8  \\ \hline
50 & 359 & 219.2  & 8 & 3.9   \\ \hline
100 & 868 & 554.5  & 8 & 4.3 \\ \hline
\end{array}
$$
\end{table}

\section{Experimental Analysis: LRP}\label{sec:lrp}
\subsection{VNS baseline algorithm}
For our comparative study, we take as a baseline algorithm a variant of the generalized VNS algorithm described in~\cite{lrpvns}. The VNS algorithm was carefully designed with LRP specific neighborhood combinations. More precisely, the baseline VNS has two standard components: random shaking and local search both using different neighborhood structures. For local search, a standard VND is used with the following neighborhoods: $\m{N}_1 \cup \m{N}_2$, $\m{N}_3 \cup \m{N}_4$, $\m{N}_5 \cup \m{N}_6$, $\m{N}_7 \cup \m{N}_8$, $\m{N}_9 \cup \m{N}_{10}$, and $\m{N}_{11}$. For shaking, both neighborhoods $\m{N}_1$, $\m{N}_2$ and $\m{N}_{11}$ are combined to produce a random neighbor each time the VND local search fails producing an improving local optimum. The maximum strength of the shaking is fixed to be a function of the problem size, namely, $n+m$. The standard VNS shaking strategy is used, i.e., if a new improving local optimum is found, shaking is rested to neighborhood $1$, otherwise the next neighborhood is considered and so on until reaching the last neighborhood. This specific shaking and local search is motivated in~\cite{lrpvns} by its ability to manage the two decision levels (Location and Routing) induced by LRP.

Generally speaking, VNS is particularly interesting for our NTS study since it uses a shaking phase combined with an efficient variable neighborhood descent. Since NTS is not equipped with any specific shaking (perturbation) procedure, the goal is to study whether the backtracking component of NTS is able to efficiently escape the local optima computed in the descent phase and to effectively find better ones without any specific shaking (perturbation). Proving that \emph{non} problem specific backtracking strategies can lead to competitive algorithms would help the design of generic search algorithms that can be applied without any problem specific tuning. In the following, starting from a randomly generated initial solution, different solution quality / computing cost trade-offs are obtained depending on NTS backtracking strategy, step function and acceptance criterion.

\subsection{Solution Quality vs Computing Cost}
We first examine solution quality obtained with NTS using FD and BD local step strategies compared to VNS. We select ordinal data analysis to compare the considered algorithms. For each algorithm $a$ and each experiment $\ell$, an ordinal value $o_a^{\ell}$ representing the rank of the algorithm is given. To compare the relative performance of competing algorithms, we aggregate the obtained orders for each algorithm into a unique order. We use a simple and intuitive aggregation method, known as the \emph{Borda count} voting method. An algorithm having rank $o_a^{\ell}$ in an experiment is given $o_a^{\ell}$ points, and the total score of an algorithm is simply the sum of its ranks over all experiments. The algorithms are then compared to their cumulative scores where the algorithm having the \emph{smallest score} being considered as the best performing algorithm. For each instance, the ranks were computed using as a metric the solution gap to lower bound averaged over $30$ trials (The lower bounds were taken from the work in~\cite{ulrp}.). In other words, for each instance, the algorithm having the $i^{\textrm{th}}$ smaller average gap is ranked $i$ and thus it is scored with $i$ points. The final score of each algorithm is them the sum of its scores over all instances. For LRP, we consider $5$ instance sets according to problem size. Each set contains $90$ instances. We will consider $7$ algorithms, i.e., $6$ NTS variants and VNS. Thus, for a given instance set, the best (resp. worst) possible score is $90$ (resp. $630$), while the best (resp. worst) possible \emph{total} score is $450$ (resp. $3150$). Our first results are summarized in Table~\ref{tab:borda}. We can observe that for lower instances sizes, FD step strategies outperforms VNS with all backtrack strategies. However, for higher instances sizes only BU performs better than VNS where as BH is the worst performing strategy overall. We attribute this to the diversification introduced by the BU strategy. Actually, the solution quality results reported in Table~\ref{tab:borda} have a price in terms of computing cost as discussed below.

\begin{table}[htb!]
\caption{Solution quality with Borda count voting method for LRP using NTS variants and VNS. Notation NTS-$(X,Y,Z)$ was defined in Section~\ref{sec:nts}. Acceptance criterion AA is fixed for all variants. $X\in\{BD,FD\}$ refers to step function. $Z\in \{ BH, BR, BU\}$ is the backtracking strategy. In bold, we highlight the scores in favor of NTS over VNS.}
\label{tab:borda}
\vspace{-3ex}
$$
\begin{array}{|c||c|c||c|c||c|c||c|}
\hline
\multirow{4}{*}{(n,m)} &\multicolumn{6}{|c||}{\textrm{NTS}-(*,AA,*)} & \multirow{4}{*}{\textrm{VNS}}\\
\cline{2-7}
 & \multicolumn{2}{|c||}{\textrm{BH}} & \multicolumn{2}{|c||}{\textrm{BR}} & \multicolumn{2}{|c||}{\textrm{BU}} & \\ \cline{2-7}

 & \textrm{BD} & \textrm{FD} & \textrm{BD} & \textrm{FD} & \textrm{BD} & \textrm{FD} &\\ 
 \cline{1-8}
 \\[-9 pt] 
\cline{1-8}

(10,30) & 598 &438 &519 &348 & \mathbf{230} & \mathbf{99} &286 \\ \hline
(10,20) &607 &471 &490 &370 & \mathbf{218} &\mathbf{115} &248 \\ \hline 
(5,30) &488 & \mathbf{359} &401 & \mathbf{266} & \mathbf{141} & \mathbf{91} &399 \\ \hline
(5,20) &452 & \mathbf{301} & \mathbf{349} &\mathbf{190} & \mathbf{131} & \mathbf{93} &403 \\ \hline
(5,10) &368 & \mathbf{250} & 322 & \mathbf{225} & \mathbf{203} & \mathbf{115} &310 \\ \hline\\[-17 pt]
& \multicolumn{2}{|c||}{ } & \multicolumn{2}{|c||}{ } & \multicolumn{2}{|c||}{ } &\\ \hline
\textrm{\textbf{Total}} &2513 &1819 &2081 & \mathbf{1399} & \mathbf{923} & \mathbf{513} &1646\\ \hline
\end{array}
$$
\end{table}

\begin{table}
\caption{Solution quality and computing cost of NTS with FD and AA strategies compared to VNS.}
\label{tab:qc}
\vspace{-3ex}
$$
\begin{array}{|c||c|c||c|c||c|c||}
\hline
\multirow{4}{*}{(n,m)} &\multicolumn{6}{|c||}{\textrm{NTS}-(FD,AA,*)} \\
\cline{2-7}
 & \multicolumn{2}{|c||}{\textrm{BH}} & \multicolumn{2}{|c||}{\textrm{BR}} & \multicolumn{2}{|c||}{\textrm{BU}} \\ \cline{2-7}

 & n_{>} & \overline{r_{\textrm{eval}}} & n_{>} & \overline{r_{\textrm{eval}}}  & n_{>} & \overline{r_{\textrm{eval}}} \\ 
 \cline{1-7}
 \\[-9 pt] 
\cline{1-7}
(10,30) & -60 & 0.50 & -37 & 0.84 & 73 &  5.00	 \\ \hline
(10,20) & -74 & 0.45 & -52 & 0.77 & 56 & 3.75	\\ \hline	
(5,30) & 20 & 0.45 & 44 & 0.78 & 70 & 4.19	 \\ \hline
(5,20) & 40 & 0.76 & 53 & 1.32 & 59 & 6.18	\\ \hline
(5,10) & 19 & 0.82 & 24 & 1.40  & 47 & 4.67 \\ \hline\\[-17 pt]
& \multicolumn{2}{|c||}{ } & \multicolumn{2}{|c||}{ } & \multicolumn{2}{|c||}{ } \\ \hline
\textrm{\textbf{Total}} & -55 & 0.60 & 32 & 1.02 & 305 & 4.76 \\ \hline
\end{array}
$$
\end{table}

In Table~\ref{tab:qc}, we report the joint solution quality, and computing cost (until termination) for NTS compared to VNS. We denote $\overline{r_{\textrm{eval}}}$ the ratio obtained when dividing the total number of evaluations performed by NTS by the total number of evaluations performed by VNS. We denote $n_{>}$ a Borda like score computed as following. For each instance, we compare the average gap to lower bound of NTS and VNS. If NTS is better we score it $+1$, in case of equality we score it $0$, and otherwise $-1$. $n_{>}$ is then obtained by summing up the computed scores. Having $90$ instances per problem size, the best (resp. worst) score is $+90$ (resp. $-90$). A positive (resp. negative, zero) score means that NTS performs better on more (resp. less, equally) number of instances. This gives a general idea on the number of instances for which NTS performs better than VNS (Taking $n_{>}/90$ gives the ratio of instances where NTS performs better or worst depending on $n_{>}$ sign). For simplicity we only report results with FD step function. Notice that Table~\ref{tab:qc} gives an idea not only about the performance of NTS compared to VNS, but also the relative performance of the different NTS strategies. One can clearly see the different trade-offs given by NTS in terms of solution quality (BH $<$ BR $\simeq$ VNS $<$ BU) and computing cost (BU $<$ VNS $\simeq$ BR $<$ BH). E.g., for lower instance sizes, BH beats VNS in both two measures. BR gives better solution quality with comparable running cost. At higher instance sizes, only BU is able to perform better than VNS in solution quality but at the price of being around $4$ times slower.

\subsection{Speeding up the search}
In this section, we give the results we have obtained by running NTS with acceptance criterion AT. Recall that with the AT strategy a neighborhood branch is explored depending on the best locally observed fitness $f_s^{\textrm{best}}$,  and with a probability which is inversely proportional to the trajectory path length. Results with FD step function are summarized in Table~\ref{tab:fdar}. One can clearly see that the AT strategy has the effect of speeding-up the search (compared to results with AA given in Table~\ref{tab:qc}). This is rather expected since neighborhood branches producing solutions with poor quality compared to other neighborhoods are likely to be pruned as we get deeper in the search path. While strategy AT allows us to speed up the search, it has two 'side-effects'. Firstly, compared to AA, AT produces less high solution quality for all backtracking strategies. Secondly and for the largest instances, AT is no more competitive against the finely tuned VNS even using the most effective BU backtracking strategy.
\vspace{-3ex}
\begin{table}[htb!]
\caption{Solution quality and computing cost of NTS with FD and AT compared to VNS.}
\label{tab:fdar}
\vspace{-3ex}
$$
\begin{array}{|c||c|c||c|c||c|c||}
\hline
\multirow{4}{*}{(n,m)} &\multicolumn{6}{|c||}{\textrm{NTS}-(FD,AT,*)} \\
\cline{2-7}
 & \multicolumn{2}{|c||}{\textrm{BH}} & \multicolumn{2}{|c||}{\textrm{BR}} & \multicolumn{2}{|c||}{\textrm{BU}} \\ \cline{2-7}

 & n_{>} & \overline{r_{\textrm{eval}}} & n_{>} & \overline{r_{\textrm{eval}}}  & n_{>} & \overline{r_{\textrm{eval}}} \\ 
 \cline{1-7}
 \\[-9 pt] 
\cline{1-7}

(10,30) & -80 & 0.22 & -66 & 0.36 & -9 & 1.03 \\ \hline
(10,20) & -88 & 0.23 & -82 & 0.35 & -38 & 0.88 \\ \hline
(5,30) & -22 & 0.22 & 13 & 0.35 & 66 & 0.92 \\ \hline
(5,20) & 16 & 0.40 & 42 & 0.63 & 55 & 1.54 \\ \hline
(5,10) & 0 & 0.53 & 14 & 0.81 & 32 & 1.53 \\ \hline\\[-17 pt]
& \multicolumn{2}{|c||}{ } & \multicolumn{2}{|c||}{ } & \multicolumn{2}{|c||}{ } \\ \hline
\textrm{\textbf{Total}} & -174 & 0.32 & -79 & 0.50 & 106 & 1.18 \\ \hline
\end{array}
$$
\end{table}

\subsection{Time to best with step function FI}

\begin{table}[htb!]
\caption{Solution quality and computing cost of NTS with FI and AA compared to VNS.}
\label{tab:fiaa}
\vspace{-3ex}
$$
\begin{array}{|c||c|c||c|c||c|c||}
\hline
\multirow{4}{*}{(n,m)} &\multicolumn{6}{|c||}{\textrm{NTS}-(FI,AA,*)} \\
\cline{2-7}
 & \multicolumn{2}{|c||}{\textrm{BH}} & \multicolumn{2}{|c||}{\textrm{BR}} & \multicolumn{2}{|c||}{\textrm{BU}} \\ \cline{2-7}

 & n_{>} & \overline{r'_{\textrm{eval}}} & n_{>} & \overline{r'_{\textrm{eval}}}  & n_{>} & \overline{r'_{\textrm{eval}}} \\ 
 \cline{1-7}
 \\[-9 pt] 
\cline{1-7}

(10,30) & -25 & 0.67 & -31 & 0.69 & -24 & 0.70  \\ \hline 
(10,20) & 44 & 1.87 & 58 & 2.14 & 67 & 2.48 \\ \hline
(5,30) & 70 & 0.26 & 70 & 0.26 & 70 & 0.26 \\ \hline 
(5,20) & 59 & 0.40 & 59 & 0.44 & 59 & 0.42  \\ \hline
(5,10) & 46 & 0.18 & 48 & 0.24 & 50 & 0.27 \\ \hline\\[-17 pt]
& \multicolumn{2}{|c||}{ } & \multicolumn{2}{|c||}{ } & \multicolumn{2}{|c||}{ } \\ \hline
\textrm{\textbf{Total}} & 194 & 0.68 & 204 & 0.75 & 222 & 0.83 \\ \hline
\end{array}
$$
\end{table}

\begin{table}[htb!]
\caption{Solution quality and computing cost of NTS with FI and AT compared to VNS.}
\label{tab:fiar}
\vspace{-3ex}
$$
\begin{array}{|c||c|c||c|c||c|c||}
\hline
\multirow{4}{*}{(n,m)} &\multicolumn{6}{|c||}{\textrm{NTS}-(FI,AT,*)} \\
\cline{2-7}
 & \multicolumn{2}{|c||}{\textrm{BH}} & \multicolumn{2}{|c||}{\textrm{BR}} & \multicolumn{2}{|c||}{\textrm{BU}} \\ \cline{2-7}

 & n_{>} & \overline{r'_{\textrm{eval}}} & n_{>} & \overline{r'_{\textrm{eval}}}  & n_{>} & \overline{r_{\textrm{eval}}} \\ 
 \cline{1-7}
 \\[-9 pt] 
\cline{1-7}

(10,30) & 11 & 1.49 & 35 & 2.64 & -27 & 0.69 \\ \hline
(10,20) & 8 & 0.84 & 33 & 1.45 & 69 & 2.47  \\ \hline
(5,30) & 70 & 0.39 & 70 & 0.48 & 70 & 0.26  \\ \hline
(5,20) & 59 & 0.34 & 59 & 0.40 & 59 & 0.44  \\ \hline
(5,10) & 42 & 0.16 & 48 & 0.19 & 50 & 0.24 \\ \hline\\[-17 pt]
& \multicolumn{2}{|c||}{ } & \multicolumn{2}{|c||}{ } & \multicolumn{2}{|c||}{ } \\ \hline
\textrm{\textbf{Total}} & 190 & 0.64 & 245 & 1.03 & 221 & 0.82 \\ \hline 
\end{array}
$$
\end{table}

In accordance with the results obtained for SMTWTP, we found that NTS combined with FI step function is able to give very good results for LRP. In the following, we report only our findings when running NTS for a maximum number of evaluations, namely, $10^7$ evaluations. For all competing algorithms, VNS included, we study the number of evaluations it takes for an algorithm to find the best fitness solution. Using acceptance conditions AA and AT, we report the values of $n_{>}$ and $\overline{r'_{\textrm{eval}}}$ which is now the ratio of the number of fitness evaluations it takes for NTS and VNS to find the best solution. Our results are summarized in Tables~\ref{tab:fiaa} and~\ref{tab:fiar}.
In accordance with the results obtained with FD, different trade-offs are obtained. For instance, the computing cost of BH is better than BR which is better than BU. For all instance sets, but for size $(10,30)$, BU beats all other strategies in terms of solution quality. Actually, for instance set $(10,30)$ strategy BU needs more time to find high quality solutions, i.e., BU is an exploration oriented strategy which needs more time to converge but produces very high solution quality. Moreover, we can state that overall instance set NTS with FI strategy performs better than VNS. In particular, backtracking strategies BH and BR provides very competitive results both in computing cost and solution quality especially when combined with the adaptive acceptance criterion AT.


\section{Conclusion}
\label{sec:conc}
In this paper, by operating at the level of the tree induced by a set of several different neighborhood structures, we introduced a backtracking traversal algorithm called NTS and studied some of its variants. Compared to standard VND where neighborhood ordering can be critical, NTS is able to find its way by simply piping different neighborhoods dynamically at runtime. Compared to VNS where shaking is crucial, backtracking in NTS is able to escape local optima searching for promising neighborhood paths. However, since exploring the neighborhood tree in an exhaustive manner could be intractable, NTS components (step function, neighborhood selection, branching, accepting, backtracking) have to be carefully combined in order to obtain a good compromise between solution quality and computing cost. In particular, the NTS variants described in this paper are based on the following two intuitive claims: (i) the more we are deep in the neighborhood tree, the more it is likely to find better local optima (intensification) (ii) the less we are deep in the neighborhood tree, the more it is likely to explore new search regions and thus to go forward through new high quality solutions (diversification). Generally speaking, we claim that new adaptive backtracking strategies combined with new adaptive acceptance criteria would be the key ingredients providing the efficient balance between intensification and diversification in NTS. We believe that this is a challenging and interesting open question which deserves further investigations.


\bibliographystyle{amsplain}

\bibliography{nts}

\providecommand{\bysame}{\leavevmode\hbox to3em{\hrulefill}\thinspace}
\providecommand{\MR}{\relax\ifhmode\unskip\space\fi MR }
\providecommand{\MRhref}[2]{%
  \href{http://www.ams.org/mathscinet-getitem?mr=#1}{#2}
}
\providecommand{\href}[2]{#2}
\begin{thebibliography}{10}

\bibitem{or}
http://people.brunel.ac.uk/~mastjjb/jeb/orlib/wtinfo.html.

\bibitem{ulrp}
M.~Albareda-Sambola, J.~A. Diaz, and E.~Fernandez, \emph{A compact model and
  tight bounds for a combined location-routing problem}, Computers \&
  Operations Research \textbf{32} (2005), no.~3, 407 -- 428.

\bibitem{ILS}
M.~D. Besten, T.~Stutzle, and M.~Dorigo, \emph{Design of iterated local search
  algorithms: An example application to the single machine total weighted
  tardiness problem}, EvoWorkshops, 2001, pp.~441--452.

\bibitem{Hyp}
E.~K. Burker, M.~Hyde, G.~Kendall, G.~Ochoa, E.~\"{O}zcan, and J.~R. Woodward,
  \emph{A classification of hyper-heuristic approaches}, Handbook of Meta.,
  2010, pp.~449--468.

\bibitem{Hyp2}
K.~Chakhlevitch and P.~I. Cowling, \emph{Hyperheuristics: Recent developments},
  Adaptive and Multilevel Metaheuristics, 2008, pp.~3--29.

\bibitem{Cow}
P.~I. Cowling, G.~Kendall, and E.~Soubeiga, \emph{A hyperheuristic approach to
  scheduling a sales summit}, $3^{\textrm{th}}$ Int. Conf. on Pract. and Th. of
  Auto. Timetabling, 2001, pp.~176--190.

\bibitem{lrpvns}
H.~Derbel, B.~Jarboui, H.~Chabchoub, S.~Hanafi, and N.~Mladenovic, \emph{A
  variable neighborhood search for the capacitated location-routing problem},
  4th IEEE Int. Conf. on Logistics (LOGISTIQUA),, 2011, pp.~514 --519.

\bibitem{galrp}
H.~Derbel, B.~Jarboui, S.~Hanafi, and H.~Chabchoub, \emph{Genetic algorithm
  with iterated local search for solving a location-routing problem}, Expert
  Syst. Appl. \textbf{39} (2012), no.~3, 2865--2871.

\bibitem{duhamel}
C.~Duhamel, Lacom P., C.~Prins, and C.~Prodhon, \emph{A grasp\&els approach for
  the capacitated location-routing problem}, Comput. Oper. Res. \textbf{37}
  (2010), 1912--1923.

\bibitem{GS}
L.~Gaspero and A.~Schaerf, \emph{Neighborhood portfolio approach for local
  search applied to timetabling problems}, J. of Mathematical Modelling and
  Algorithms \textbf{5} (2006), 65--89.

\bibitem{vnd}
Martin~Josef Geiger, \emph{On heuristic search for the single machine total
  weighted tardiness problem - some theoretical insights and their empirical
  verification}, EJOR \textbf{207} (2010), no.~3, 1235--1243.

\bibitem{GMK}
A.~Goeffon, J.-M. Richer, and J.-K. Hao, \emph{Progressive tree neighborhood
  applied to the maximum parsimony problem}, IEEE/ACM Trans. Comput. Biol. Bio.
  \textbf{5} (2008), 136--145.

\bibitem{dyna}
A.~Grosso, F.~Della Croce, and R.~Tadei, \emph{An enhanced dynasearch
  neighborhood for the single-machine total weighted tardiness scheduling
  problem}, Operations Research Letters \textbf{32} (2004), no.~1, 68 -- 72.

\bibitem{VNS}
P.~Hansen, N.~Mladenovi\'c, and  P.~J. Moreno, \emph{Variable neighbourhood
  search: methods and applications}, Annals of Operations Research \textbf{175}
  (2010), 367--407.

\bibitem{HS}
H.~H. Hoos and T.~St\"{u}tzle, \emph{Evaluating las vegas algorithms: pitfalls
  and remedies}, $14^{\textrm{th}}$ Conf. on Uncertainty in artificial
  intelligence, 1998, pp.~238--245.

\bibitem{HR}
B.~Hu and G.~Raidl, \emph{Variable neighborhood descent with self-adaptive
  neighborhood-ordering}, $7^{\textrm{th}}$ EU/MEeting on Adaptive,
  Self-Adaptive, and Multi-Level Meta., 2006.

\bibitem{LGH}
Z.~L\"{u}, F.~Glover, and J.-K. Hao, \emph{Neighborhood combination for
  unconstrained binary quadratic problems}, MIC Post-Conference Book, 2011,
  pp.~49--61.

\bibitem{LKG}
Z.~L\"{u}, J.-K. Hao, and F.~Glover, \emph{Neighborhood analysis: a case study
  on curriculum-based course timetabling}, J. of Heuristics \textbf{17} (2011),
  97--118.

\bibitem{Min}
H.~Min, V.~Jayaraman, and R.~Srivastava, \emph{Combined location-routing
  problems: A synthesis and future research directions}, EJOR \textbf{108}
  (1998), no.~1, 1 -- 15.

\bibitem{lrp}
G.~Nagy and S.~Salhi, \emph{Location-routing: Issues, models and methods}, EJOR
  \textbf{177} (2007), no.~2, 649--672.

\bibitem{Ozcan}
E.~\"{O}zcan, B.~Bilgin, and E.~E. Korkmaz, \emph{A comprehensive analysis of
  hyper-heuristics}, Intell. Data Anal. \textbf{12} (2008), 3--23.

\bibitem{PR}
J.~Puchinger and G.~Raidl, \emph{Bringing order into the neighborhoods:
  relaxation guided variable neighborhood search}, J. of Heuristics \textbf{14}
  (2008), 457--472.

\end{thebibliography}

\balancecolumns

\end{document}